\documentclass[conference]{IEEEtran}
\IEEEoverridecommandlockouts
\usepackage{cite}
\usepackage{amsmath,amssymb,amsfonts}
\usepackage{algorithmic}
\usepackage{graphicx}
\usepackage{stfloats}
\usepackage{textcomp}
\usepackage{subfig}
\usepackage{xcolor}
\usepackage{booktabs}
\usepackage{multirow}
\def\BibTeX{{\rm B\kern-.05em{\sc i\kern-.025em b}\kern-.08em
    T\kern-.1667em\lower.7ex\hbox{E}\kern-.125emX}}
\begin{document}

\title{Beyond KNN: Deep Neighborhood Learning for WiFi-based Indoor Positioning Systems\\

}

\author{\IEEEauthorblockN{
1\textsuperscript{st} Yinhuan Dong, 
2\textsuperscript{nd} Francisco Zampella,
3\textsuperscript{rd} Firas Alsehly}
\IEEEauthorblockA{\textit{Edinburgh Research Centre} \\
\textit{Huawei Technologies Research and Development (UK) Ltd}\\
Edinburgh, UK \\
\{yinhuan.dong1, francisco.zampella, firas.alsehly\}@huawei.com}
%\and
%\IEEEauthorblockN{}
%\IEEEauthorblockA{\textit{Edinburgh Research Centre} \\
%\textit{Huawei Technologies Research and Development (UK) Ltd}\\
%%Edinburgh, UK \\
%francisco.zampella@huawei.com}
%\and
%\IEEEauthorblockN{3\textsuperscript{rd} Firas Alsehly}
%\IEEEauthorblockA{\textit{Edinburgh Research Centre} \\
%\textit{Huawei Technologies Research and Development (UK) Ltd}\\
%Edinburgh, UK \\
%firas.alsehly@huawei.com}

}

\maketitle

\begin{abstract}
K-Neares Neighbors (KNN) and its variant weighted KNN (WKNN) have been explored for years in both academy and industry to provide stable and reliable performance in WiFi-based indoor positioning systems. Such algorithms estimate the location of a given point based on the locality information from the selected nearest WiFi neighbors according to some distance metrics calculated from the combination of WiFi received signal strength (RSS). However, such a process does not consider the relational information among the given point, WiFi neighbors, and the WiFi access points (WAPs). Therefore, this study proposes a novel Deep Neighborhood Learning (DNL). The proposed DNL approach converts the WiFi neighborhood to heterogeneous graphs, and utilizes deep graph learning to extract better representation of the WiFi neighborhood to improve the positioning accuracy. Experiments on 3 real industrial datasets collected from 3 mega shopping malls on 26 floors have shown that the proposed approach can reduce the mean absolute positioning error by 10\% to 50\% in most of the cases. Specially, the proposed approach sharply reduces the root mean squared positioning error and 95\% percentile positioning error, being more robust to the outliers than conventional KNN and WKNN.     
\end{abstract}


\section{Introduction}
The proliferation of smartphones has provided an excellent tool to keep people connected, provide them with the latest information, and in general making life easier for many. These devices and the applications around them provide services tailored to the user and their context, but they rely on up to date information about the device. Among the most useful information is the position of the user, that in outdoor scenarios is typically obtained using Global Navigation Satellite Systems. In indoor scenarios due to the attenuation and multipaths of the signals, the accuracy of the positioning is severely affected \cite{van1996multipath} and the service needs to rely on other technologies and techniques to improve it.

In most indoor environments, the devices are able to connect to many wireless networks that can be used for positioning including cellular, WiFi, Bluetooth, Ultra Wide Band, among others. The most widely deployed technology is WiFi, and as an example, in shopping malls it is typically possible to observe between ten and hundreds of WiFi access points (WAP) at any given point near shops. The WiFi technology offers several metrics that can be used for positioning \cite{liu2020}, including the angle of Arrival \cite{gallo2015, wen2014}, time of arrival \cite{golden2007, danjo2015}, time difference of arrival \cite{nawaz2017} and the received signal strength (RSS) \cite{hernandez2017, ding2013}. Among those, the RSS has become one of the most commonly used sources of information because it does not require extra hardware and it is available in all devices. In open spaces, it is possible to implement a trilateration from the position of the WAPs, and estimating the distances to the user with the RSS measured in dB and a log distance model \cite{hidayab2009}, but even if the positions were known, that model changes according to the unknown conditions of the obstructions between the user and the WAPs.

As a way to overcome the non predictable distribution of the RSS values with position, is it common to measure the RSS values from all observable WAPs in a grid of samples covering all the areas of interest. These measurements can be used to estimate more complex models of the RSS-distance models \cite{atia2012}, the RSS-position model \cite{wang2020}, as reference points (fingerprints) in a weighted K-Nearest Neighbors (WKNN) estimation \cite{torres2015, Manhattan1, Manhattan2, Manhattan3}, or other machine learning methods \cite{feng2014, zhang2014, chen2018, shenoy2019}. In most cases it has been observed that the wide distribution of the data in the RSS-distance and RSS-position modeling doesn't allow to have a local model, but in the WKNN estimation, it only use the locality of the neighbors, but we loose the information from the RSS values. The authors of this paper propose \emph{Deep Neighborhood Learning} to analyse the RSS distribution using a deep graph neural network-based model but focusing in the local community or neighborhood of similar reference points.
% to get better representation of the WiFi FP to improve the positioning accuracy
% Furthermore, such algorithms make predictions by simply taking mean or weighted mean of the selected neighbors, which is not robust enough to provide accurate estimations. To overcome the challenges of missing information from the RSS and provide better learning from the WiFi neighborhood, the author....    

\par The contributions of this work are summarized as follows:
\begin{itemize}
\item We propose a novel Deep Neighborhood Learning approach to extract better representation of the WiFi neighborhood to improve the positioning accuracy. 
\item In the proposed approach, we develop a deep graph neural network-based positioning model to learn the relational information (the topology and the features) from the WiFi neighborhood.
\item Compared to conventional positioning algorithms using the WiFi neighborhood, the experiments on three real large datasets  have shown that the proposed approach can sharply reduce the positioning error and the influence from the outliers.
\end{itemize}

The rest of the paper is distributed as follows: Section \ref{sec_methodology} will discuss the proposed architecture, Section \ref{sec_experiment_setup} will describe the experiments, Section \ref{sec_results} will show the results and analyse them, and in Section \ref{sec_conclusions} we will draw the general conclusions.

%\section{Related work}

\section{Methodology}\label{sec_methodology}
As illustrated above, to overcome the challenges of missing RSS features in conventional KNN and its variant WKNN, we propose a \emph{Deep Neighborhood Learning} approach to learn the relation among one given fingerprint (FP) and its WiFi neighbors, and hence provide better positioning accuracy for WiFi-based indoor positioning systems.

\subsection{The local community}
A local community represents a snapshot of the RSS distributions in the vicinity of any given FP. Each local community is constructed from a center/target FP (TFP) and its closest neighbor FPs (NFP), as well as all the WiFi measurements that can be detected among the FPs in the local community. 
\par To select the NFPs of a certain TFP, we first calculate the distance from the center/target FP to all the reference FPs. In this work, we use the Manhattan distance to represent the distance between two FPs in the signal space. The Manhattan distance has been shown to be effective to represent the similarity between two WiFi fingerprints, and has been adopted by many WiFi fingerprinting-based positioning algorithms \cite{Manhattan1, Manhattan2, Manhattan3}. Supposing there are $n$ WAPs in the entire dataset, the RSS values of one FP are converted to a vector of length $n$, where the non-detected signals are padding with a default value of 0. So that the Manhattan distance between the two FP vectors of $FP_1 = [RSS_1^1, RSS_2^1, ... RSS_n^1]$ and $FP_2 = [RSS_1^2, RSS_2^2, ... RSS_n^2]$ can be calculated by:

\begin{equation}
|FP_1-FP_2|=\sum_{i=1}^n\left|RSS_i^1-RSS_i^2\right|
\end{equation}
Therefore, for $m$ FPs, we will have $m$ TFPs, and accordingly $m-1$ distances from a certain TFP to all other reference FPs. We select the top $k$ FPs with the shortest distance as the NFPs. The local community will consist of the target FP and the neighbor FPs as:
\begin{equation}
C = \{TFP, NFP_1, NFP_2,..., NFP_{k}\}
\end{equation}

\subsection{Graph representations of local communities}
\begin{figure}
\centering
\includegraphics[width=0.5\textwidth]{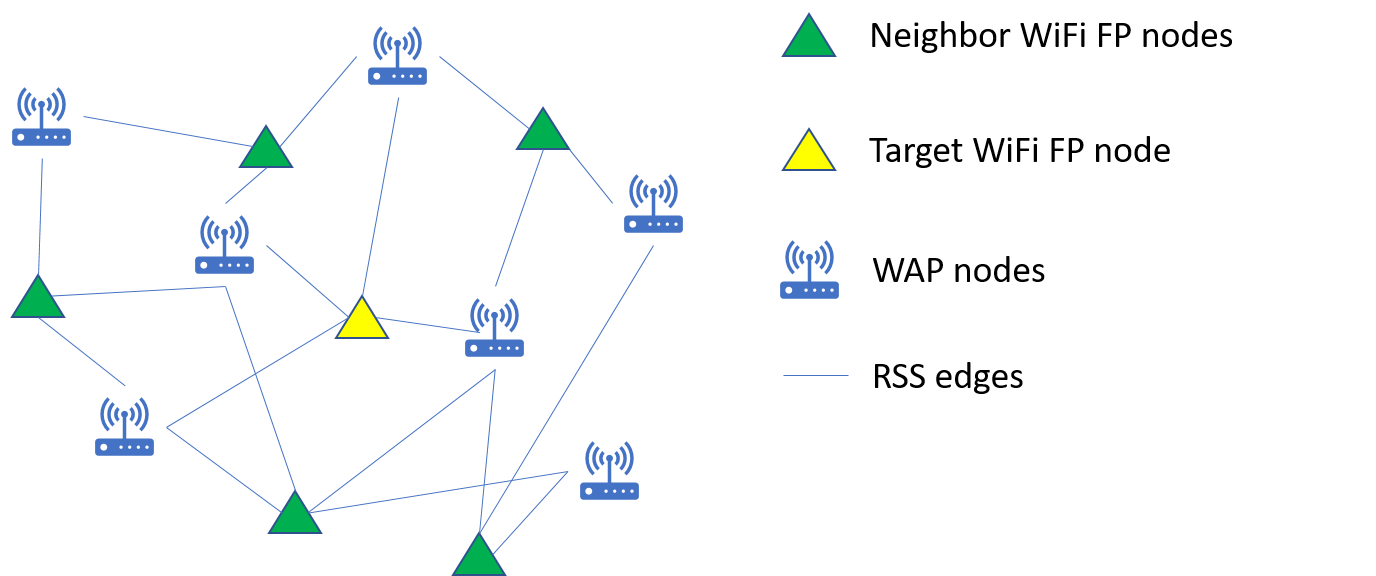}
\caption{\label{subgrapha}An example of the graph representation of the local community for training.}
\end{figure}

The local communities represent more than just the fingerprints that are similar to the TFP, the relationships established by the RSS observed by the members of that set to the WAPs in the vicinity offer a glimpse of the local model of RSS in there. The community can be represented as a heterogeneous graph as described in \cite{zhang2022}, but focusing only in the area of interest. 
%Now we translate the local communities to heterogeneous graphs for later positioning use. 
Given a training reference set and a test set, 2 types of graphs constructions are required, the neighbors graphs in the training reference set, where each reference FP finds neighbors in the rest of the training set, and the neighbors graphs in the test set, where each test FP finds the neighbors among the training set.
%First, we consider the graph construction of the samples in the dataset for training.
In the training case, the graph shown in Figure \ref{subgrapha} can be created with 3 types of nodes, the neighbor FPs, the target FP and the WAP nodes. The neighbor FP nodes include information about the position of that reference point ($F_W$), the target FP has a position masked with zeros, and the WAP node stores the MAC index of the WAP ($F_A$). The observations of each FP ($FP_i=\{MAC_{i_1}:RSS_{i_1}, ..., MAC_{i_n}:RSS_{i_n}\}$) are encoded in the edges between FP nodes and WAP nodes ($E_{W \rightarrow A}$ and $E_{A \rightarrow W}$), where the edge from (to) WAP node $i$ to (from) FP node $j$ has a weight equal to the RSS observed in FP $j$ for WAP $i$.
%As one example of the graph shown in Figure \ref{subgrapha},  we have two types of nodes: WiFi FP nodes $W$ and AP nodes $A$. We set the locations of where the WiFi FPs were taken as the node features $F_W$ of all NFP nodes; while the node feature of the TFP node is masked with an all zero vector. The AP node features $F_A$ are represented by consecutive integer indices from 0 to $n-1$.  The connections between such nodes are based on the observations of the RSS values. Therefore, there are two type of edges: $E_{W \rightarrow A}$ and $E_{A \rightarrow W}$. Both of the edges are weighted by the RSS values. 
Each graph of the local community of TFP $i$ in the dataset for training can be denoted by $g_{train}(i) = \{F_W, F_A, E_{W \rightarrow A}, E_{A \rightarrow W} \}$. Each graph is associated with a label of the location of the TFP node.

\begin{figure}
\centering
\includegraphics[width=0.5\textwidth]{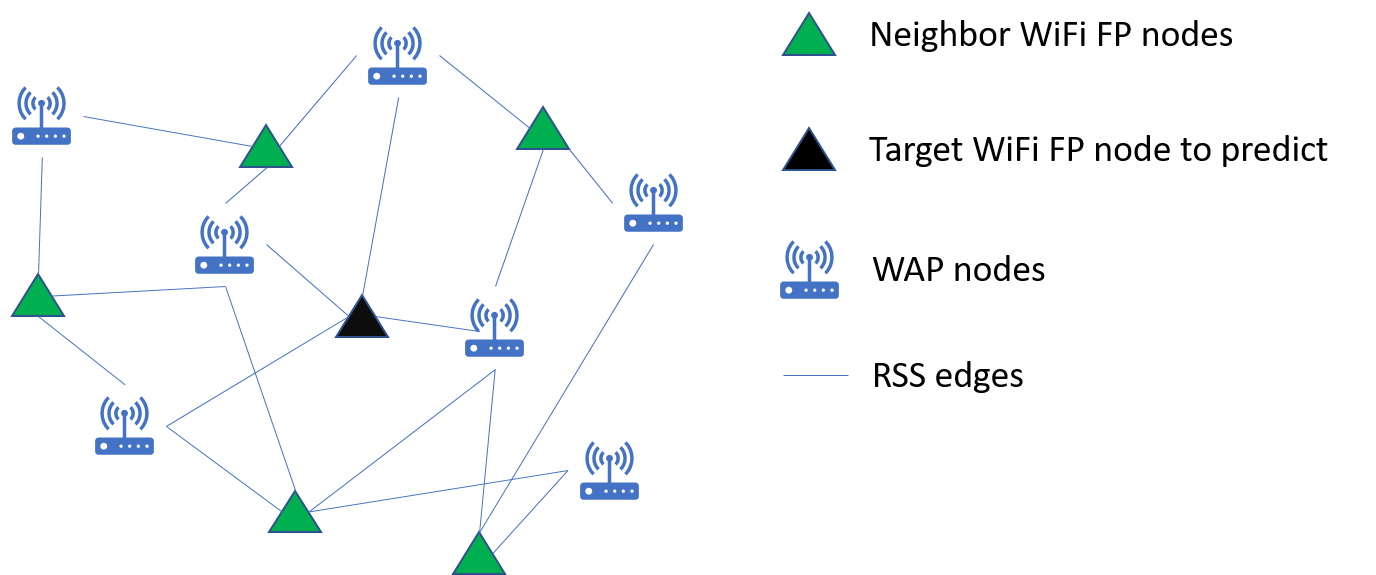}
\caption{\label{subgraphb}An example of the graph representation of the local community for inference/testing.}
\end{figure}

For inference of the test set, the TFP $j$ has no label, but the same method to construct a local community is used, selecting the NFPs from the training dataset, and constructing a similar graph $g_{inf}(j)$ as shown in Figure \ref{subgraphb}.

\subsection{Positioning with Deep Neighborhood Learning}
\begin{figure*}
\centering

\includegraphics[width=0.9\textwidth]{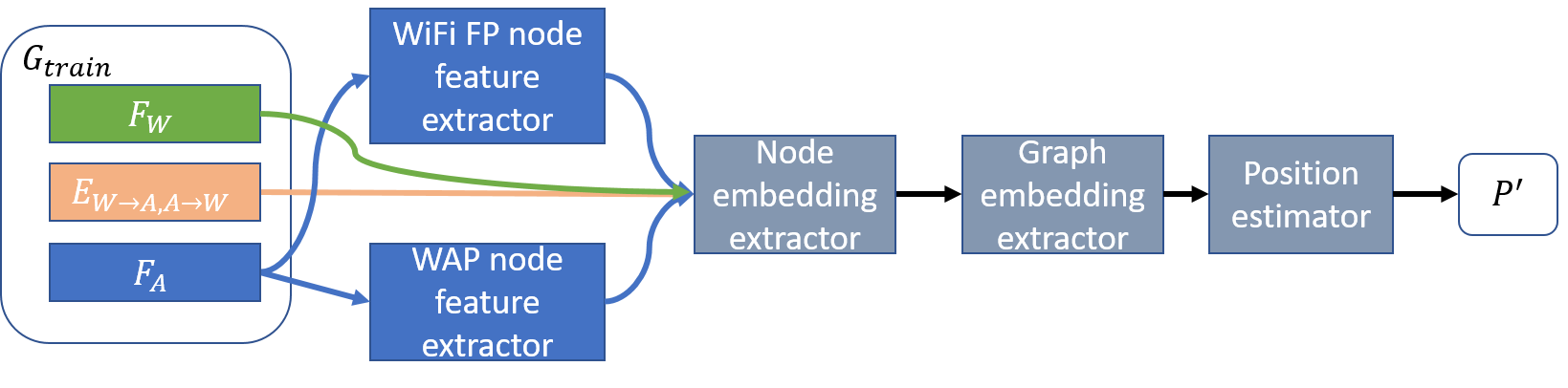}
\caption{Structure of the proposed supervised deep graph neural networks-based positioning model.}
\label{network}
\end{figure*}
\par The positioning problem is now translated to making inference of the unlabeled graph $g_{inf}$ by a model trained with labeled graphs $G_{train} = \{g_{1},g_{2},...,g_{n}\}$. The labels are denoted by $P = {p_1, p_2, ... p_n}$, where $p_i$ denotes the position of the $i$th $g_{train}$. To solve such a problem, we propose a supervised model based on graph neural networks. As the structure of the proposed model is shown in Figure \ref{network}, the model is constructed of multiple modules:
\begin{itemize}
    \item WiFi FP feature extractor: this module is a 3-layer perceptron with hidden size of 8 to extract latent features from $F_W$.
    \item WAP feature extractor: this module is constructed of one embedding layer with hidden size of 8 to extract the embedding of the Mac address indices $F_A$.
    \item Node embedding extractor: this module contains two Graph Isomorphism Network (GIN) \cite{GIN} layers. It aggregate the node features from both types of the nodes through the weighted edges (edges with edge features) and extract the node embedding.
    \item Graph embedding extractor: this module takes the sum of the mean readout of each type of nodes as the graph embedding.
    \item Position estimator: this module is a 2-layer perceptron with hidden size of 64 to make predictions from the graph embedding.
\end{itemize}
\par To train the model, we minimize the loss of mean squared error (MSE) between the labels $P$ and the predictions $P'$. The MSE loss function is defined as follows:
\begin{equation}
\mathrm{MSE}=\frac{1}{b} \sum_{i=1}^b\left(P_i-{P}'\right)^2
\end{equation}
where $b$ denotes the number of training samples in each batch.

\section{Experimental Settings} \label{sec_experiment_setup}
\subsection{Data}
The proposed method was evaluated on three large Huawei WiFi indoor positioning datasets from: Joycity, Universal Harbor, and Dawan Mall (three huge shopping malls in Beijing and Shanghai in China). The datasets are obtained from users submitted data, calibrated using the method described in \cite{gu2018} and manually map matched to the floor plans.
%\begin{itemize}
%    \item Joycity: Shopping with 10 shopping floors and 3 underground floors in %Beijing,
    %\item Universal harbor:
    %\item Dawan mall:
%\end{itemize}

The details of the datasets are shown in Tables \ref{joycity}, \ref{universalharbor} and \ref{dawanmall}.
\begin{table*}[]
\centering
\caption{Details of the Huawei indoor positioning dataset I (Joycity)}
\label{joycity}
\begin{tabular}{lllllllllllllll}
\hline
Floor                         &  -3  & -2   & -1   & 1     & 2    & 3    & 4    & 5    & 6    & 7    & 8    & 9 &10   & Total \\ \hline
Number of FPs       & 1188 & 2215 & 3953 & 30386 & 8834 & 3386 & 3132 & 5310 & 6240 & 3354 & 1912 & 1446 &1827 & 70168 \\
% Number of  GT graphs              & 482  & 584  & 1056  & 1343 & 730  & 738  & 661  & 780  & 505  & 409  & 364  & 7652  \\
Detected Mac addresses       &1165     & 2112 & 3941 & 5675  & 4381 & 3111 & 2562 & 2440 & 2752 & 2085 & 1445 & 900 &835  & 8971  \\
% Date when the data were collected &      &      &       &      &      &      &      &      &      &      &      &       \\
\hline
\end{tabular}
\end{table*}

\begin{table*}[]
\centering
\caption{Details of the Huawei indoor positioning dataset II (Universal harbor)}
\label{universalharbor}
\begin{tabular}{llllllllll}
\hline
Floor                  & -3   & -2    & -1    & 1     & 2    & 3    & 4    & 5    & Total  \\ \hline
Number of FPs       & 1681 & 18484 & 17491 & 38763 & 6044 & 7641 & 8676 & 2136 & 100916 \\
Detected Mac addresses & 1271 & 3196  & 3646  & 5992  & 3834 & 2608 & 2387 & 1891 & 8384\\  
\hline
\end{tabular}
\end{table*}

\begin{table*}[]
\centering
\caption{Details of the Huawei indoor positioning dataset III (Dawan mall)}
\label{dawanmall}
\begin{tabular}{lllllll}
\hline
Floor                  & -3   & -2    & -1   & 1     & 2    & Total \\ \hline
Number of FPs       & 4144 & 41121 & 1640 & 19784 & 7078 & 73767 \\
Detected Mac addresses & 1279 & 2348  & 2766 & 6366  & 3623 & 8728  \\ \hline
\end{tabular}
\end{table*}

\subsection{Training settings}
\par Each dataset (per floor) was partitioned into training, validation and test sets according to the ratio of 6:2:2. In this study, we focus on 2-dimension positioning problem, and hence we trained floor-based model for each floor in each building. Each model maintained the same structure. 
\par For better fitting of the model, we supervised the variation of the validation loss between two epochs and adjust the learning rate dynamically. Initially, the learning rate was set to 0.01. The learning rate was reduced by a factor of 0.1 if the validation loss does not decrease in 3 epochs till the learning rate reaches the minimum of 0.0001. 
\par Besides, we conducted repeated training with different batch size (64, 128, and 256) of graphs in each training session of 100 epochs. In each training session, the model with the lowest validation loss was saved for later testing and evaluations.

\subsection{Evaluation Metrics}
To evaluate the positioning accuracy, we calculated the mean absolute error (MAE) and the root mean squared error (RMSE):
\begin{equation}
\begin{aligned}
\text { RMSE } &=\sqrt{\frac{\sum_{i=1}^n e_i}{n}} \\
\text { MAE } &=\frac{1}{n} \sum_{i=1}^n \sqrt e_i\\
\end{aligned}
\end{equation}
where:
\begin{equation}
e_i  = \left(P_{i,x}-{P'}_{i,x}\right)^2 +\left(P_{i,y}-{P'}_{i,y}\right)^2 
\end{equation}
\par Also, we compared the cumulative distribution error by sorting the $e_i$ from lowest to highest and take the square root of the 68\% and 95\% point, respectively.

\section{Result and Analysis}\label{sec_results}
\begin{figure*}[!t]
\centering
\subfloat[]{\includegraphics[width=0.32\textwidth]{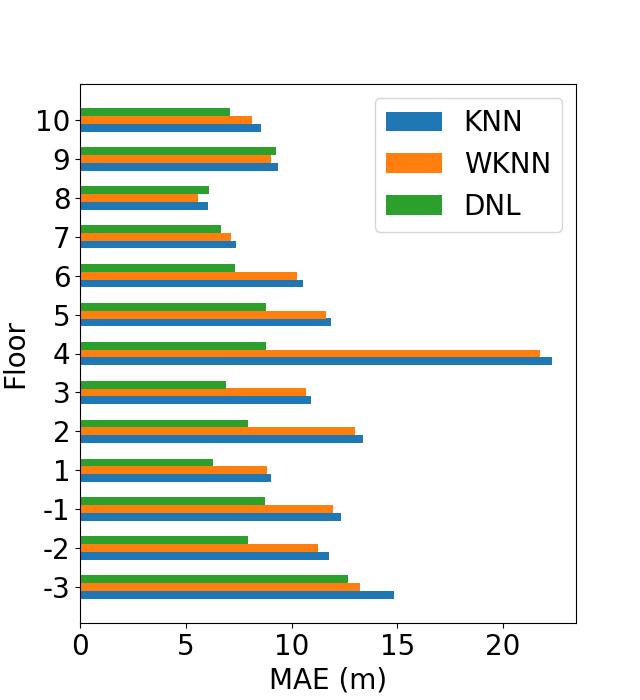}
\label{joy_city_c42MAE}}
\hfil
\subfloat[]{\includegraphics[width=0.32\textwidth]{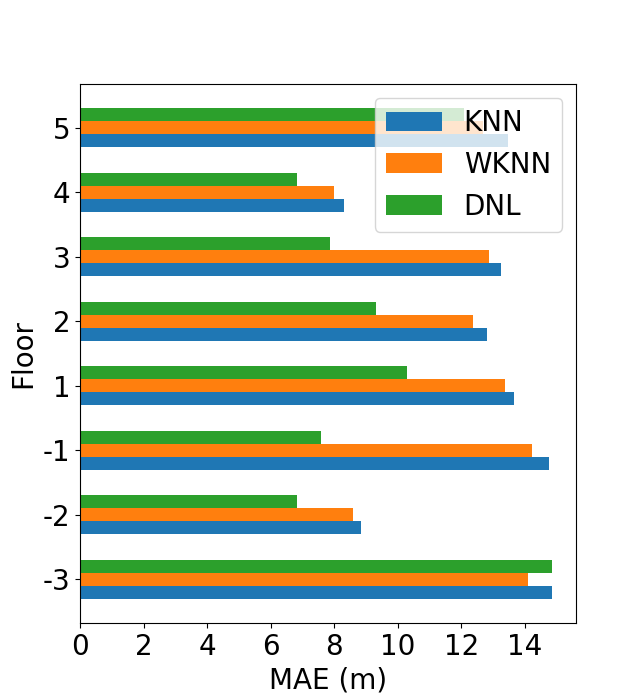}
\label{universal_harbor_c45MAE}}
\hfil
\subfloat[]{\includegraphics[width=0.32 \textwidth]{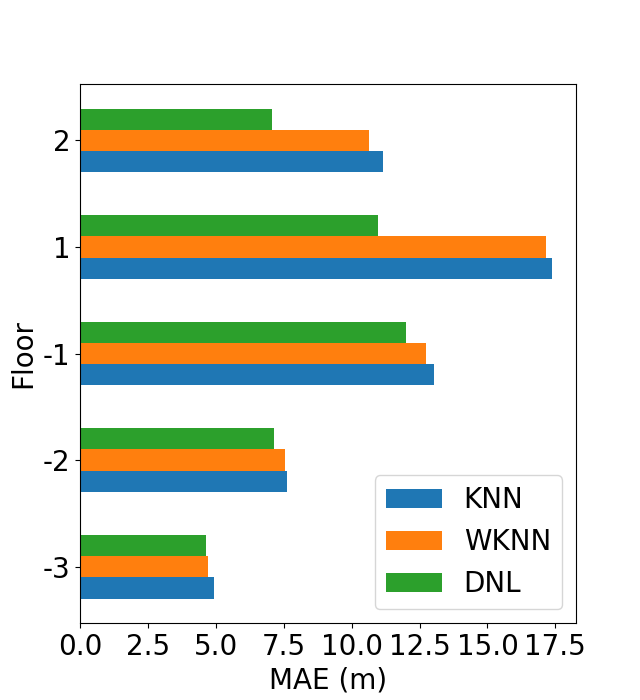}
\label{dawan_mall_c35MAE}}
\caption
{Comparisons of mean absolute error using different positioning algorithms on the three buildings: (a) Joycity; (b) Universal Harbor; (c) Dawan Mall.}
\label{MAE}
\end{figure*}

\begin{figure*}[!t]
\centering
\subfloat[]{\includegraphics[width=0.32\textwidth]{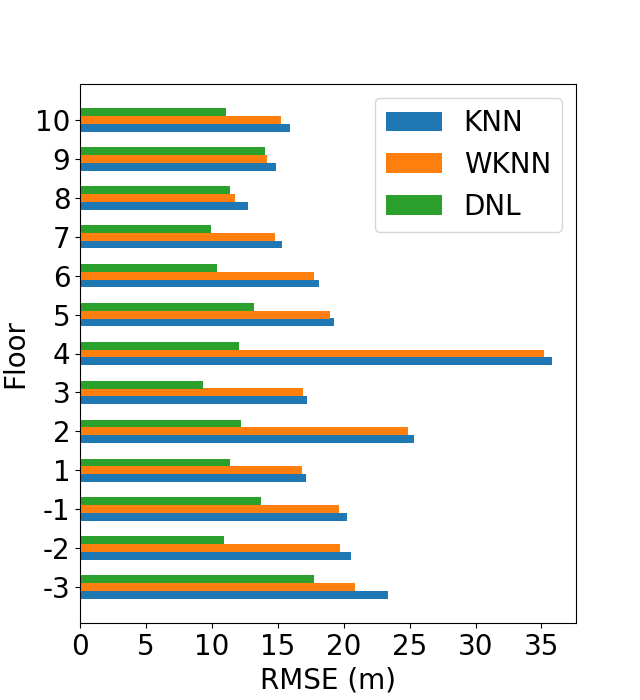}
\label{joy_city_c42MAE}}
\hfil
\subfloat[]{\includegraphics[width=0.32\textwidth]{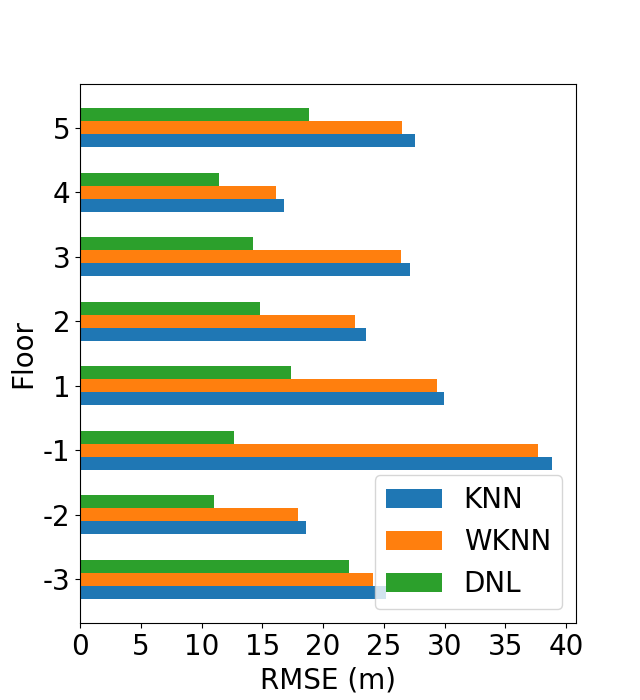}
\label{universal_harbor_c45MAE}}
\hfil
\subfloat[]{\includegraphics[width=0.32 \textwidth]{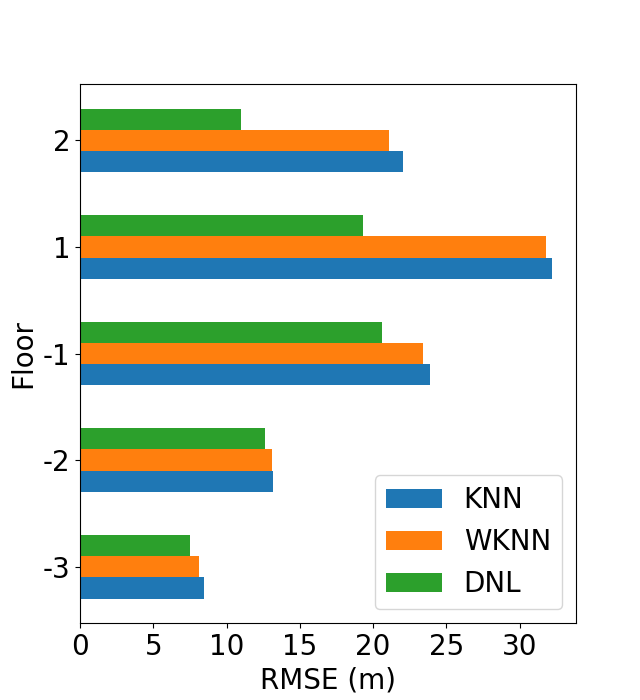}
\label{dawan_mall_c35MAE}}
\caption
{Comparisons of root mean squared error using different positioning algorithms on the three buildings: (a) Joycity; (b) Universal Harbor; (c) Dawan Mall.}
\label{RMSE}
\end{figure*}

In this section, we analyze the experimental results and evaluate the proposed model. Specially, We compare the positioning results to an unsupervised learning algorithm of k-nearest neighbors (KNN) which uses the similar strategy of WiFi neighbor selection. As illustrated previously, KNN and its variant Weighted KNN (WKNN) calculate the position of one given test point by taking the mean or the weighted mean of the selected WiFi neighbors. Such algorithms narrow down the positioning problem from the entire interested area (floor, in this study) to a local community constructed by the neighbor WiFi fingerprints. However, they do not have a learning process to infer the final location but simply calculate the mean or weighted mean of the neighbors location, which attributes to low inference ability.  

\par Two standard positioning algorithms of KNN and WKNN were implemented for comparison in this study:
\begin{itemize}
    \item KNN: implemented with scikit-learn K Nearest Neighbors Regression \cite{KNN}; number of neighbor was set to 10; Manhattan distance was set to the distance metric. 
    \item WKNN: similar to KNN but also set the weight as the inverse the distance in prediction. 
\end{itemize}

\par As shown in Figure \ref{MAE}, using the MAE, the proposed model outperform the KNN and WKNN in 25 and 23 out of 26 cases. In detail, we can see from Table \ref{cases} that the proposed model can provide a 10\% to 50\% reduction of the MAE in 18 (KNN) and 16 (WKNN) of the cases. Rather than simply taking the mean or the weighted mean position of the WiFi neighbors, the proposed model can well learn the WiFi neighborhood through the node features, edge features, and the topology information to make better predictions of the location.
\par The proposed model shows even better improvement in RMSE. As we can observe from the Table \ref{cases}, the proposed model shows lower RMSE than the other two algorithms in all 26 cases. Particularly, the reduction of RMSE in most of the cases (19 cases against KNN; 15 cases against WKNN) is higher than 30\%. Additionally, the proposed model provides much lower RMSE in some specific cases, particularly when the dataset has a large data volume and a large amount of detected Mac addresses. For instance, 5992 Mac addresses have been detected among 38763 FPs on the first floor of Universal harbor. Such large dataset usually contains more uncertainty and noise; however, the proposed model is able to provide approximatelly 40\% less RMSE than the other two algorithms. Similar cases can also be found in Joycity and Dawan Mall from Table \ref{allerror} and \ref{RMSE}. This illustrates that the proposed model can provide even much smoother and more accurate predictions than the other two if the data is noisier.     
\par Last bu not least, the cumulative positioning error of different algorithms was compared. It can be seen from Table \ref{allerror} that the proposed model challenged KNN and WKNN in half of the cases, when using the 68 percentile of the positioning accuracy (68\% of the CDF). In the remaining cases, it can be observed that the improvement of the proposed model degrade when the cases have lower data density. For example, on floor -3 and 5 in Universal Harbor, the proposed model is 18\% and 28\% worse than WKNN, respectively. Similar results can also be found in other floors, particularly some top floors and underground floors with less data. Nevertheless, the proposed model shows again superb performance in the 95 percentile positioning accuracy (95\% of the CDF). The model can always provide much lower error than KNN and WKNN, which suggest that the proposed algorithm is more robust to outliers in large datasets like the ones observed in the middle floors or in lower density areas like the top or underground floors.
\begin{table*}[]
\centering
\caption{Number of cases that the proposed model shows better positioning performance than KNN and WKNN}
\label{cases}
\begin{tabular}{@{}ccccccccc@{}}
\toprule
\multirow{2}{*}{\begin{tabular}[c]{@{}c@{}}Percentage of \\ improvement\end{tabular}} & \multicolumn{2}{c}{MAE} & \multicolumn{2}{c}{RMSE}& \multicolumn{2}{c}{68\%CDF}& \multicolumn{2}{c}{95\%CDF} \\ \cmidrule(l){2-9}  & KNN        & WKNN       & KNN        & WKNN  & KNN        & WKNN & KNN        & WKNN       \\ \midrule
0-10\%                                                                       & 6          & 6          & 2          & 5 
                    & 9          & 9          & 0          & 5\\
10\% - 30\%                                                                           & 9          & 10         & 5          & 6                        & 7          & 5          & 9          & 3\\
30\% - 50\%                                                                           & 9          & 6          & 15         & 12                       & 0          & 0          & 7          & 8 \\
\textgreater{}50\%                                                                    & 1          & 1          & 4          & 3                        & 1          & 1          & 10          & 9\\ \midrule
Total                                                                                 & 25         & 23         & 26         & 26
                    & 17         & 15         & 26         & 25\\ \bottomrule
\end{tabular}
\end{table*}

% \begin{figure*}[!t]
% \centering
% \subfloat[]{\includegraphics[width=0.32\textwidth]{results_cep/knn.png}
% \label{knncdf}}
% \hfil
% \subfloat[]{\includegraphics[width=0.32\textwidth]{results_cep/wknn.png}
% \label{wknncdf}}
% \hfil
% \subfloat[]{\includegraphics[width=0.32 \textwidth]{results_cep/gnn.png}
% \label{gnncdf}}
% \caption
% {Comparisons of the cumulative error using different positioning algorithms on all the floors: (a) KNN; (b) WKNN; (c) GNN.}
% \label{cdf}
% \end{figure*}

% \usepackage{multirow}
\begin{table*}[]
\centering
\caption{Comparison of the positioning error using different algorithms}
\label{allerror}
\begin{tabular}{llllllllllllll}
\hline
\multirow{2}{*}{Building}         & \multirow{2}{*}{Floor} & \multicolumn{3}{l}{MAE}                         & \multicolumn{3}{l}{RMSE}       & \multicolumn{3}{l}{68\%CDF}             & \multicolumn{3}{l}{95\%CDF}             \\ \cline{3-14} 
                                  &                        & KNN           & WKNN           & DNL            & KNN   & WKNN  & DNL            & KNN   & WKNN           & DNL            & KNN   & WKNN           & DNL            \\ \hline
Joycity                            & -3                     & 14.84         & 13.24          & \textbf{12.67} & 23.37 & 20.88 & \textbf{17.72} & 15.47 & 14.41          & \textbf{14.05} & 48.65 & 43.19          & \textbf{39.00} \\
                                  & -2                     & 11.76         & 11.25          & \textbf{7.94}  & 20.54 & 19.71 & \textbf{10.94} & 8.39  & 8.20           & \textbf{8.01}  & 47.76 & 45.30          & \textbf{22.04} \\
                                  & -1                     & 12.33         & 11.96          & \textbf{8.76}  & 20.23 & 19.65 & \textbf{13.76} & 10.59 & 9.95           & \textbf{8.44}  & 46.93 & 46.22          & \textbf{27.00} \\
                                  & 1                      & 9.05          & 8.84           & \textbf{6.26}  & 17.15 & 16.81 & \textbf{11.36} & 6.76  & 6.68           & \textbf{5.59}  & 35.43 & 34.77          & \textbf{17.70} \\
                                  & 2                      & 13.36         & 13.00          & \textbf{7.92}  & 25.33 & 24.85 & \textbf{12.22} & 8.61  & 8.39           & \textbf{7.46}  & 62.57 & 60.31          & \textbf{24.49} \\
                                  & 3                      & 10.91         & 10.67          & \textbf{6.90}  & 17.24 & 16.91 & \textbf{9.36}  & 10.00 & 9.78           & \textbf{7.32}  & 40.60 & 39.99          & \textbf{18.50} \\
                                  & 4                      & 22.33         & 21.74          & \textbf{8.80}  & 35.83 & 35.17 & \textbf{12.07} & 21.97 & 20.44          & \textbf{9.43}  & 85.42 & 85.35          & \textbf{26.23} \\
                                  & 5                      & 11.87         & 11.61          & \textbf{8.78}  & 19.27 & 18.96 & \textbf{13.17} & 10.76 & 10.19          & \textbf{9.19}  & 43.42 & 42.96          & \textbf{22.20} \\
                                  & 6                      & 10.55         & 10.27          & \textbf{7.30}  & 18.12 & 17.74 & \textbf{10.41} & 8.07  & 7.78           & \textbf{7.48}  & 45.08 & 43.90          & \textbf{19.62} \\
                                  & 7                      & 7.37          & 7.14           & \textbf{6.68}  & 15.31 & 14.80 & \textbf{9.90}  & 5.93  & \textbf{5.76}  & 6.57           & 23.56 & 23.04          & \textbf{21.08} \\
                                  & 8                      & \textbf{6.06} & 5.59           & 6.08           & 12.72 & 11.73 & \textbf{11.38} & 4.87  & \textbf{4.78}  & 5.60           & 20.24 & \textbf{16.07} & 17.13          \\
                                  & 9                      & 9.35          & \textbf{9.04}  & 9.28           & 14.85 & 14.22 & \textbf{14.00} & 9.02  & 8.82           & \textbf{8.79}  & 34.09 & 32.95          & \textbf{29.69} \\
                                  & 10                     & 8.54          & 8.13           & \textbf{7.09}  & 15.96 & 15.24 & \textbf{11.05} & 6.53  & \textbf{6.16}  & 6.33           & 37.33 & 36.14          & \textbf{26.24} \\ \hline
\multirow{8}{*}{Universal Harbor} & -3                     & 14.86         & \textbf{14.09} & 14.84          & 25.20 & 24.08 & \textbf{22.14} & 12.70 & \textbf{12.26} & 14.87          & 59.26 & 53.66          & \textbf{49.46} \\
                                  & -2                     & 8.83          & 8.59           & \textbf{6.84}  & 18.55 & 17.93 & \textbf{11.06} & 6.23  & \textbf{6.13}  & 6.54           & 34.90 & 33.02          & \textbf{20.22} \\
                                  & -1                     & 14.76         & 14.22          & \textbf{7.60}  & 38.86 & 37.69 & \textbf{12.65} & 6.73  & \textbf{6.61}  & 7.11           & 77.68 & 75.46          & \textbf{22.59} \\
                                  & 1                      & 13.65         & 13.37          & \textbf{10.29} & 29.91 & 29.33 & \textbf{17.34} & 9.37  & \textbf{9.23}  & 9.24           & 56.25 & 54.77          & \textbf{32.10} \\
                                  & 2                      & 12.80         & 12.37          & \textbf{9.31}  & 23.49 & 22.59 & \textbf{14.78} & 10.31 & 9.95           & \textbf{9.17}  & 53.98 & 52.28          & \textbf{28.85} \\
                                  & 3                      & 13.26         & 12.86          & \textbf{7.88}  & 27.14 & 26.41 & \textbf{14.23} & 7.54  & 7.38           & \textbf{7.22}  & 69.00 & 67.91          & \textbf{22.98} \\
                                  & 4                      & 8.31          & 8.00           & \textbf{6.84}  & 16.74 & 16.11 & \textbf{11.42} & 6.42  & \textbf{6.28}  & 6.69           & 30.84 & 27.90          & \textbf{17.60} \\
                                  & 5                      & 13.47         & 12.69          & \textbf{12.10} & 27.55 & 26.53 & \textbf{18.85} & 9.60  & \textbf{8.57}  & 11.98          & 65.35 & 63.82          & \textbf{44.16} \\ \hline
\multirow{5}{*}{Dawan Mall}       & -3                     & 4.95          & 4.73           & \textbf{4.64}  & 8.44  & 8.09  & \textbf{7.49}  & 4.42  & \textbf{4.35}  & 4.44           & 17.05 & 15.12          & \textbf{12.54} \\
                                  & -2                     & 7.62          & 7.56           & \textbf{7.14}  & 13.17 & 13.10 & \textbf{12.65} & 7.09  & 7.03           & \textbf{6.69}  & 20.87 & 20.51          & \textbf{18.48} \\
                                  & -1                     & 13.04         & 12.75          & \textbf{12.02} & 23.89 & 23.39 & \textbf{20.61} & 9.24  & \textbf{9.04}  & 10.70          & 62.01 & 61.03          & \textbf{47.32} \\
                                  & 1                      & 17.39         & 17.15          & \textbf{10.99} & 32.23 & 31.81 & \textbf{19.30} & 12.07 & 11.96          & \textbf{9.49}  & 78.74 & 78.30          & \textbf{37.01} \\
                                  & 2                      & 11.14         & 10.63          & \textbf{7.07}  & 22.02 & 21.07 & \textbf{10.97} & 7.36  & 7.09           & \textbf{6.70}  & 52.87 & 51.13          & \textbf{23.25} \\ \hline
\end{tabular}
\end{table*}

\section{Conclusions}\label{sec_conclusions}
In conclusion, we have proposed a Deep Neighborhood Learning approach to compensate the intrinsic problem of losing relational information among FPs and WAPs in conventional KNN or WKNN-based WiFi fingerprinting algorithms. For a given target FP, the proposed approach constructs a heterogeneous graph based on its local community containing the WiFi neighbor FPs and all the WAPs that can be detected within the community. A special neural network model has been designed to learn the topology information and the features from the neighborhood in the graph, and project the extracted embedding to the location of the target FP. Compared to KNN and WKNN, experiments on 3 real industrial datasets collected from 3 shopping malls, covering 26 floors in total, have shown that the proposed approach can better learn the neighborhood information and reduce the positioning error (MAE and RMSE) by 10\% to 50\% in most of the cases. Particularly, the DNL approach can always provide much lower RMSE and 95\% percentile positioning error, being more robust to the outliers caused by the noise from huge datasets. 
\par Our future work will be conducted from two aspects. On one hand, we will integrate the WiFi distance learning model proposed in our recent work to select better WiFi neighbors; On the other hand, we will include more information to the local community to extract even better representation of the WiFi fingerprints, such as the measurements and calculations from the inertial sensors and GNSS data.  

\section*{Acknowledgment}

This research is from a internship project within Huawei Technologies Research and Development (UK) Ltd Edinburgh Research Centre. We thank our colleagues Bertrand Perrat, Ilari Vallivaara, Janis Sokolovskis and Rory Hughes for their great work on data mining and post-processing.

\end{document}